\documentclass[twocolumn,secnumarabic,superscriptaddress,amssymb,aps,prl,10pt,showpacs]{revtex4-1}
\usepackage{natbib}
\usepackage[USenglish]{babel}
\usepackage{amsmath,amssymb,amsxtra,mathtools,bbm}
\usepackage[normalem]{ulem}
\usepackage{titlesec}
\usepackage{filecontents}
\usepackage{color}
\usepackage{upgreek}
\usepackage[breaklinks]{hyperref}

\hypersetup{
    unicode=false,          
    pdftoolbar=true,        
    pdfmenubar=true,        
    pdffitwindow=false,     
    pdfstartview={FitH},    
    pdfnewwindow=true,      
    colorlinks=true,       
    linkcolor=blue,          
    citecolor=blue,        
    filecolor=blue,      
    urlcolor=blue           
}

\newcommand{\eqr}[1]{Eq.~(\ref{#1})}

\newcommand{\fir}[1]{Fig.~\ref{#1}}
\newcommand{\Fir}[1]{Fig.~\ref{#1}}
\newcommand{\secr}[1]{Sec.~\ref{#1}}

\DeclareMathOperator{\diag}{diag}

\DeclareMathOperator{\Tr}{Tr}

\newcommand{\half}{\mbox{$\textstyle \frac{1}{2}$}}

\newcommand{\expt}[1]{\langle #1 \rangle}
\newcommand{\commutator}[2]{\left[#1,#2\right]}

\newcommand{\mum}{$\upmu$m}

\titleclass{\prlsection}{straight}[\section]

\titleformat{\prlsection}[runin]
  {\itshape\normalsize}{}{0em}{}[.---]
\titlespacing*{\prlsection}{\parindent}{0ex}{0.0\parindent}

\makeatletter
  \def\toclevel@prlsection{1}
  \def\l@prlsection{\@dottedtocline{3}{3.8em}{3.2em}}
\makeatother

\newenvironment{widetextnoline@grid}{%
  \par\ignorespaces
  \setbox\widetext@top\vbox{%
   \hb@xt@\hsize{%
    \leaders\hrule\hfil
    \vrule\@height6\p@
   }%
  }%
  \setbox\widetext@bot\hb@xt@\hsize{%
    \vrule\@depth6\p@
    \leaders\hrule\hfil
  }%
  \onecolumngrid
  \vskip10\p@
  \dimen@\ht\widetext@top\advance\dimen@\dp\widetext@top
  \cleaders\box\widetext@top\vskip\dimen@
  \vskip6\p@
  \prep@math@patch
}{%
  \par
  \vskip6\p@
  \setbox\widetext@bot\vbox{%
   \hb@xt@\hsize{\hfil\box\widetext@bot}%
  }%
  \dimen@\ht\widetext@bot\advance\dimen@\dp\widetext@bot
  \cleaders\box\widetext@bot\vskip\dimen@
  \vskip8.5\p@
  \twocolumngrid\global\@ignoretrue
  \@endpetrue
}%

\definecolor{highlight}{rgb}{1,0,0}

\allowdisplaybreaks

\begin{document}

\title{\texorpdfstring{Proposed parametric cooling of bilayer cuprate superconductors\\ by terahertz excitation}
{Proposed parametric cooling of bilayer cuprate superconductors by terahertz excitation}}

\author{S.~J.~Denny}
\email{s.denny@physics.ox.ac.uk}
\affiliation{Clarendon Laboratory, University of Oxford, Parks Road, Oxford OX1 3PU, United Kingdom}
\author{S.~R.~Clark}
\affiliation{Clarendon Laboratory, University of Oxford, Parks Road, Oxford OX1 3PU, United Kingdom}
\affiliation{Max Planck Institute for the Structure and Dynamics of Matter, Hamburg, Germany}
\author{Y.~Laplace}
\affiliation{Max Planck Institute for the Structure and Dynamics of Matter, Hamburg, Germany}
\author{A.~Cavalleri}
\affiliation{Max Planck Institute for the Structure and Dynamics of Matter, Hamburg, Germany}
\affiliation{Clarendon Laboratory, University of Oxford, Parks Road, Oxford OX1 3PU, United Kingdom}
\author{D.~Jaksch}
\affiliation{Clarendon Laboratory, University of Oxford, Parks Road, Oxford OX1 3PU, United Kingdom}
\affiliation{Centre for Quantum Technologies, National University of Singapore, 3 Science Drive 2, Singapore 117543}

\date{\today}

\begin{abstract}
We propose and analyze a scheme for parametrically cooling bilayer cuprates based on the selective driving of a $c$-axis vibrational mode. The scheme exploits the vibration as a transducer making the Josephson plasma frequencies time-dependent. We show how modulation at the difference frequency between the intra- and interbilayer plasmon substantially suppresses interbilayer phase fluctuations, responsible for switching $c$-axis transport from a superconducting to resistive state. Our calculations indicate that this may provide a viable mechanism for stabilizing non-equilibrium superconductivity even above $T_c$, provided a finite pair density survives between the bilayers out of equilibrium.

\end{abstract}


\maketitle

\prlsection{Introduction}
The ability to use light to drive with precision a single low-energy degree of freedom of a solid is rapidly becoming an important tool for both basic research and potential technological applications \cite{Subedi2014,Rini2007,Forst2014,Caviglia2012,Fausti2011,Kaiser2014,Hu2014}. Much work in this area has been dedicated to the excitation of lattice vibrations, which deform the crystal lattice when 
driven to large amplitudes \cite{Forst2011}. Often, these vibrational modes lie in the mid-infrared region. The laser excitation of these modes can be coherent, highly selective and induce little direct heating, in contrast to near-visible wavelengths \cite{Miyano1997,Cavalleri2001,Iwai2002,Chollet2005,Perfetti2006}. Accordingly, such nonlinear phononic techniques have been directed toward materials with strong electronic correlations, with the goal to optically switch their collective properties including superconductivity, ferroelectricity, or colossal magnetoresistance \cite{Fausti2011,Kaiser2014}. 

Notably, the application of nonlinear lattice control in high-$T_c$ superconducting cuprates has lead to the realisation of light enhanced superconductivity, demonstrated first by targeting modes which dynamically ``unbuckle" the crystallographic structure of the cuprate La$_{1.675}$Eu$_{0.2}$Sr$_{0.125}$CuO$_4$, tipping the system from striped to superconducting behaviour \cite{Forst2014,Fausti2011}. In a recent experiment, coherent excitation of apical oxygen distortions in the bilayer cuprate YBa$_2$Cu$_3$O$_{6+d}$ (YBCO) was shown to induce a transient phase which exhibited superconducting fluctuations at temperatures up to 300K \cite{Hu2014,Kaiser2014}. In these experiments, the disruption of a competing order cannot fully explain the extraordinary temperature scale of the effect, and other phenomena related to the nature of the dynamically driven state should be considered. Consequently, here we explore the properties of bilayer cuprates under periodic driving, which in other materials systems has been shown to give rise to a renormalized electronic structure \cite{Wang2013}.
   
While the physics of high-$T_c$ superconductors is not fully understood it is generally considered that their properties are determined by the doped copper-oxide planes \cite{Pavarini2001,Weber2010,Slezak2008,Mori2008a,Sakakibara2010}. As these planes are weakly coupled through insulating layers in the $c$-axis, the low-energy $c$-axis electrodynamics of cuprates can often be adequately described as a stack of \emph{intrinsic} Josephson junctions -- making them potentially compact sources of coherent continuous-wave THz frequency radiation \cite{Hu2010,Welp2013,Bulaevskii2007}. Owing to the relatively small $c$-axis phase stiffness and poor screening phase fluctuations in the copper oxide planes are likely to play a significant role in determining the critical temperature $T_c$ \cite{Emery1995,Mihlin2009}. In this Letter we propose a cooling scheme for bilayer cuprates, similar to the laser cooling of solids via anti-Stokes fluorescence \cite{Nemova2010}, but specifically targeting the crucial order-parameter phase. 

\begin{figure}[t]
\centering
\includegraphics{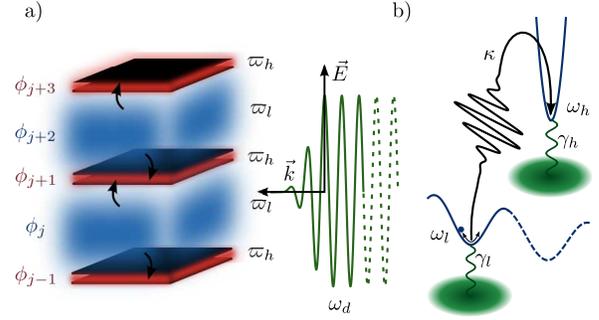}
\caption{(a)~A schematic of a bilayer cuprate such as YBCO composed of Josephson junctions each with a phase difference $\phi_j$ and alternating interbilayer (low) $\varpi_l$ and intrabilayer (high) $\varpi_h$ plasma frequencies. (b) Parametric cooling scheme where the coupling between the low and high frequency normal modes $\omega_l$ and $\omega_h$ are modulated. Tuning the modulation to $\omega_d = \half(\omega_h - \omega_l)$ THz cools the low frequency mode by up-converting fluctuations to the high frequency one.}
\label{fig:schematic}
\end{figure}

Cooling of phase fluctuations by driving has been demonstrated in a BCS superconductor using microwave frequency sideband techniques, leading to an increase in the critical current $I_c$ for a single Josephson junction \cite{Hammer2011}. Bilayer cuprates shown schematically in \fir{fig:schematic}(a), are composed of unit cells with two junctions whose insulators alternate between thick interbilayers and thin intrabilayers. Typically they have low ($l$) and high ($h$) Josephson plasma normal modes in the region $\omega_l \approx 2\pi \times 1$~THz and $\omega_h \approx 2\pi \times 10$~THz, respectively. Our key idea is to use the selective driving of a $c$-axis vibrational mode as a transducer to modulate plasma frequencies in time. Although this driving differs from that used to laser cool many-body systems of atoms~\cite{Wieman1999,Griessner2006,Jaksch2001}, ions~\cite{Eschner2003} and optomechanical oscillators~\cite{Aspelmeyer2013}, the resulting effect is analogous. Given a temperature $\hbar\omega_l < k_BT < \hbar\omega_h$, parametric modulation \cite{Vyatchanin1977} of the bilayer structure can cool by up-converting energy from the thermally populated interbilayer plasmon modes -- responsible for phase fluctuations -- to the intrabilayer plasmon modes, see \fir{fig:schematic}(b). We make testable quantitative predictions on the efficiency of cooling and the resulting elevations in $I_c$. Importantly we find that optimal suppression of phase fluctuations occurs for modulation at the frequency difference of the plasmon modes. While the theory discussed here is inspired by the experiments reported in Refs.~\cite{Kaiser2014,Hu2014}, optimal conditions were not met in these experiments so the theory outlined may or may not explain those observations. 

\prlsection{Model}
The $c$-axis electrodynamics of cuprate materials are commonly modeled as alternating stacks of superconducting and insulating layers, with the Josephson effect and quasiparticle tunnelling providing coupling along the $c$-axis \cite{Kleiner1994b, Hu2010}. The application of Maxwell's equations, augmented by the Josephson relations, yields a model in which there is both inductive and capacitive coupling between the phases of each intrinsic junction \cite{Koyama1996}. To simplify our treatment, we consider a sufficiently small crystal in the $a$ and $b$ dimensions ($< 100$ \mum{} for many cuprates) such that plasmon modes with finite quasi-momentum in the $ab$ plane are energetically prohibited. Consequently the spatial dependence of the phases can be neglected, reducing the system to a stack of short junctions, similar to those fabricated in heterostructures, with layer charging providing the dominant coupling \cite{Koyama1996}.

\begin{figure}[tb]
\centering
\includegraphics{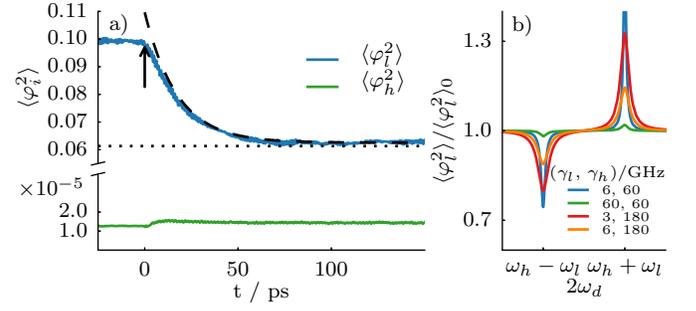}
\caption{Parametric cooling in a two-junction unit cell with (undriven) normal modes $\omega_l = 2\pi \times 1$~THz,  $\omega_h = 2\pi \times 10$~THz,  $\gamma_l = 0.19$~THz, $\gamma_h = 3.63$~THz, $\alpha = 1$, $\eta_l = 0$ and $\eta_h = 0.1$. (a)~Phase and conjugate momentum quadratures of the inter- and intrabilayer modes. 
Before $t=0$ (indicated) $T_\text{initial}/T_0 = 0.1$, then parametric driving is applied, after which fluctuations on both quadratures of the inter- (intra-) bilayer mode are cooled to 0.6$T_\text{initial}$ (heated to $1.2T_\text{initial}$). Analytic predictions of steady state (dotted line) and asymptotic cooling rate (dashed line) are also shown. (b)~Cooling/heating sidebands of the interbilayer mode as the driving frequency $\omega_d$ is varied.} 
\label{fig:two_junctions_phase_cooling}
\end{figure}

In addition to the Josephson dynamics there are also phonon modes spanning the THz range \cite{Kress1988,Prade1989,Kulkarni1989,Kulkarni1991a}, many of which describe $c$-axis vibrations within the insulating layers \cite{Liu1988}. 
The direct coupling of an infrared active vibrational coordinate $q$ to the Josephson plasma frequencies is central to the physics of this work. Such a coupling, which will be $q^2$ due to symmetry, might arise in numerous ways. For example, the motion of apical oxygens may modulate superfluid density and the plasma frequency. Alternatively, lattice vibrations may be modeled as a time-dependent modulation of the capacitance of the insulating layers, equivalent to a $\chi^{(3)}$ optical nonlinearity \cite{Boyd2003}. Either type of modulation results in the same general effect, and for concreteness we focus on the latter mechanism \cite{SupMat}. This is equivalent to the insulating layer $I$ having an effective time-dependent relative permittivity $\epsilon_I(t)$ modulated at twice the driving frequency $\omega_d$, along with off-resonant harmonics which we neglect. In the Supplemental Material \cite{SupMat} we show that the Josephson phase dynamics is described by a modified Koyama-Tachiki model \cite{Koyama1996},
\begin{multline}\label{eq:Josephson_stack_EoM}
  \partial_t^2 \phi_I = -\alpha \varpi_{I-1}^2 \sin(\phi_{I-1}) +  \left[2\alpha+ 1 + \eta_I(t)\right]\varpi_I^2\sin(\phi_I) \\
    - ~\alpha\varpi_{I+1}^2 \sin(\phi_{I+1}),
\end{multline}
where $\phi_I$ is the gauge-invariant phase difference across the $I^\text{th}$ insulating layer, $\varpi_I = c/\sqrt{\epsilon_I}\lambda_{c}$ is the layers' alternating plasma frequency given by the static permittivity $\epsilon_I$, the superconducting $c$-axis penetration depth $\lambda_{c}$ and the speed of light in vacuum $c$. The capacitive coupling between junctions is quantified by the parameter $\alpha$ which takes values in the range 0.1--5 for common high-$T_\text{c}$ superconductors \cite{Tsvetkov1998,Machida2004a}. The relative driving strength is $\eta_I(t) = (2\alpha + 1) \Lambda_I [\epsilon_I/\epsilon_I(t) - 1] = \eta_I\sin^2(\omega_d t)$, where the factor $\Lambda_I < 1$ accounts for the enlarged effective thickness of a layer due to $\lambda_c$ \cite{SupMat}. The driving therefore modulates the bare plasma frequency $\varpi_I$. An alternative approach, making use of the Lawrence-Doniach model, yields an identical equation of motion \cite{Bulaevskii1994,Hu2010}.

\prlsection{Two-junction unit cell} 
To examine the phase dynamics of a unit cell we linearize \eqr{eq:Josephson_stack_EoM} and move to the normal mode frame of \eqr{eq:Josephson_stack_EoM} with $\eta_I(t)$ replaced by its time-average $\half \eta_I$. We adopt a classical Langevin framework for describing the noise $\xi_{l,h}(t)$ and damping $\gamma_{l,h}$ caused e.g.~by long-wavelength phonons and incoherent quasi-particle currents. This gives
\begin{equation}\label{eq:normal_mode_eom}
  \begin{gathered}
  \partial_t^2 \varphi_l - \gamma_l \partial_t \varphi_l + \omega_l^2(t) \varphi_l + \Delta_h(t) \varphi_h = \xi_l(t), \\
  \partial_t^2 \varphi_h - \gamma_h \partial_t \varphi_h + \omega_h^2(t) \varphi_h + \Delta_l(t) \varphi_l = \xi_h(t),
  \end{gathered}
\end{equation}
where $\varphi_{l,h}$ are the normal mode phase coordinates. Importantly the driving introduces a time-dependent off-diagonal coupling $\Delta_{l,h}(t) = \Delta_{l,h}\cos(2\omega_d t)$, where $\Delta_{l} = -\half\Theta(\alpha)\eta_h\varpi^2_l$, using $\Theta(\alpha) = \alpha^2/(2\alpha+1)^2$, and $\Delta_h = -\half\eta_h\varpi^2_h$ to lowest order in $r = \varpi_l/\varpi_h$. Similarly the driving also induces a modulation of the normal mode frequencies $\omega_{l,h}^2(t) = \omega^2_{l,h} - \half A^2_{l,h}\cos(2\omega_d t)$, with $A^2_{l} = [\eta_l + \Theta(\alpha)\eta_h]\varpi_l^2$ and $A^2_{h} = \eta_h\varpi_h^2$. The quadratic nature of the driving shifts the normal mode frequencies $\omega_{l,h}$, in line with experimental observations \cite{Hu2014,Kaiser2014}, as explicitly shown in the Supplemental Material \cite{SupMat}. The noise $\xi_{l,h}(t)$ is approximated as independent, white and Gaussian, $\expt{\xi_{l,h}(t)\xi_{l,h}(t')} = \Gamma_{l,h} \delta(t-t')$ and is related to the damping by the fluctuation-dissipation theorem as $\Gamma_{l} = 2\gamma_{l}\omega_l^2(T/T_0)$ and $\Gamma_h = 2\gamma_h \omega_l^2(\omega_{l}/\omega_{h})^2(T/T_0)$, where $T_0$ is a system dependent temperature scale set by the capacitive energy associated to the mode $\varphi_l$. In the absence of driving the damping and noise will thermalize the system at a temperature $T$. In contrast to well isolated quantum optical/atomic systems, their continued presence during driving accounts for persistent reheating expected in a solid-state system.

We integrate the stochastic differential equations \eqr{eq:normal_mode_eom} with a quasi-symplectic velocity Verlet propagator \cite{Melchionna2007}. In \fir{fig:two_junctions_phase_cooling} we report results for a representative set of relevant parameters for bilayer cuprates when driving at the difference frequency $\omega_d = \half(\omega_h - \omega_l)$. Since the intrabilayer junction is typically more highly damped we have taken $\gamma_h > \gamma_l$. As shown in \fir{fig:two_junctions_phase_cooling}(a), once the driving is switched on the phase fluctuations of the interbilayer mode $\varphi_l$ are strongly suppressed. Up-conversion correspondingly causes fluctuations to increase on the intrabilayer mode $\varphi_h$, however its fluctuations remain small even in the driven steady state. Number fluctuations (not shown) for the two modes behave similarly. Although the resulting steady state is non-thermal the level of fluctuations is consistent with $\varphi_l$ being substantially cooled, and $\varphi_h$ being heated. The $\omega_d$ dependence of the effect is shown in \fir{fig:two_junctions_phase_cooling}(b) where a cooling (red) and heating (blue) sidebands are observed at $\omega_d = \half(\omega_h - \omega_l)$ and $\omega_d = \half(\omega_h + \omega_l)$, respectively.

\prlsection{Parametric cooling}
\begin{figure}[tb]
\centering
\includegraphics{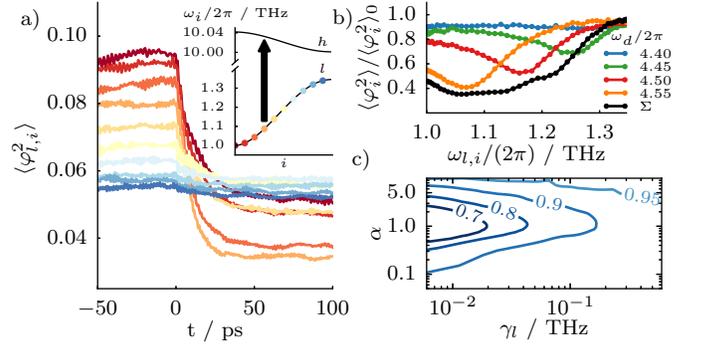}
\caption{(a)~Reduction of $\varphi_{l,i}$ fluctuations for a 100-junction stack over time, at $T/T_c \approx 0.7$. A selection of 10 modes have been displayed across the interbilayer band as shown in the inset. (b)~In the steady-state for relative fluctuations for each mode in the interbilayer band is plotted for a selection of $\omega_d$. (c)~A contour plot of the $\expt{\sin^2 \varphi_{l,i}}/ \expt{\sin^2 \varphi_{l,i}}_0$ averaged across the interbilayer band sweeping over the damping $\gamma_l$ and capacitive coupling $\alpha$. Here we have taken $\gamma_h = 0.1\min(\omega_{h,i})$, $\min(\omega_{h,i})/\min(\omega_{l,i}) = 10$. 
For each value of $\alpha$, we drive at a frequency targeting the bottom of the $l$ band.}
\label{fig:stack_cooling}
\end{figure}  
We estimate the final temperature and cooling rate by neglecting the modulation of the normal mode frequencies $\omega_m(t)$, and retaining only the modulation of the couplings $\Delta_m(t)$. The resulting model is an effective coupled-oscillator Hamiltonian, as depicted in \fir{fig:schematic}(b) 
\begin{multline}\label{eq:effective_twoJJ_H}
 H = \frac{p_l^2}{2\Delta_l}
     + \frac{p_h^2}{2\Delta_h}
     + \half \Delta_l \omega_l^2 \varphi_l^2
     + \half \Delta_h \omega_h^2 \varphi_h^2 \\
     + \cos(2\omega_d t) \Delta_l \Delta_h \varphi_l \varphi_h,
\end{multline}
where $p_{l,h}$ is the conjugate momentum to $\varphi_{l,h}$. Applying the rotating wave approximation and transforming to the frame rotating with the coupling reveals that $\omega_d = \half(\omega_h - \omega_l)$ modulation induces resonant exchange energy between the oscillators \cite{SupMat}. Since both oscillators are coupled to the same thermal reservoir up-conversion of energy from $\varphi_l$ to $\varphi_h$ is the dominant process. At the temperatures of interest the high frequency bath modes thermalising $\varphi_h$ are effectively unoccupied, so excess upconverted energy is dissipated. This leads to cooling controlled by the normalized coupling $\kappa_0^2 = (\Delta_{h}\Delta_{l})/(\omega_l \omega_h) = \frac{1}{4}(\eta_h^2 \omega_h^2)g(\alpha)\,r+ \mathcal{O}(r^3)$, with $g(\alpha) =\Theta(\alpha)/\sqrt{3 \alpha ^2+4 \alpha +1}$, and is maximized at $\alpha = \alpha_0 \approx 1.07$, where $g(\alpha_0) \approx 0.04$. Note that $\kappa_0$ depends only on $\eta_h$ to leading order in $r$, because the $\varphi_h$ mode is more massive by a ratio $(\omega_h/\omega_l)^2$, so modulation of the intrabilayer insulator is predicted to be most effective. In the Supplemental Material \cite{SupMat} we show that the asymptotic cooling rate is $\gamma_\text{dr} = \gamma_l + \kappa_0^2/(\gamma_h - \gamma_l) + \mathcal{O}(\kappa_0^3)$, and the steady state fluctuations for the interbilayer mode are \cite{Vyatchanin1977}
\begin{equation}
\expt{\varphi_l^2}/\expt{\varphi_l^2}_0 = 1 - \mathcal{S}(1- \omega_l/\omega_h),
\end{equation}
where the scale factor $0 \leq \mathcal{S} \leq 1$ is given by $\mathcal{S} = \zeta/\gamma_l \chi$, with $\zeta = \kappa_0^2 (\gamma_l + \gamma_h)/[\Delta \omega^2 + (\gamma_l + \gamma_h)^2]$, $\chi = 1 + \zeta(\gamma_l + \gamma_h)/\gamma_l \gamma_h$, and $\Delta \omega = 2\omega_d - (\omega_h - \omega_l)$. Note that $\mathcal{S}=0$, indicating no cooling, if either $\gamma_h = 0$ or $\kappa_0 = 0$, while for an undamped interbilayer mode $\gamma_l =0$ with $\gamma_h,\kappa_0>0$, we have $\mathcal{S}=1$ giving the maximum suppression of fluctuations. With increasing $\gamma_l>0$, $\mathcal{S}$ decreases monotonically implying the interbilayer should be underdamped, and $\mathcal{S}$ displays the expected resonance around $\Delta\omega = 0$. The predictions of this analysis are included in \fir{fig:two_junctions_phase_cooling}(a), and agree with the numerical solution to within a few percent over a wide parameter regime.
Moreover, in the Supplemental Material \cite{SupMat} we show that the neglect of quantum fluctuations in either plasmon mode does not affect the validity of our approach.


\prlsection{Cooling a stack of junctions}
We now turn to the main result of this Letter and consider the full nonlinear dynamics of a stack of 100 junctions described by \eqr{eq:Josephson_stack_EoM} using the classical Langevin treatment outlined. As shown in \fir{fig:stack_cooling}(a)(inset) the linearized normal modes of the stack now form bands, $\varphi_{l,i}$ and $\varphi_{h,i}$, of low and high frequency plasmons which we take as being uniformly damped at rates $\gamma_l$ and $\gamma_h$, respectively. The $c$-axis critical temperature $T_c$ of the stack was determining by locating when spontaneous thermal phase slipping first induces a resistive state under negligible bias \cite{SupMat}. 
In \fir{fig:stack_cooling}(a) we show the fluctuations of the modes $\varphi_{l,i}$ for the case where the driving $\omega_d$ is tuned to half the difference frequency near the lower edge of the bands (indicated).
As with the two junction case a suppression of phase fluctuations is observed for nearby modes. The steady-state driven fluctuations of $\varphi_{l,i}$ when $\omega_d$ targets different regions of the interbilayer band are shown in \fir{fig:stack_cooling}(b). Also plotted is the response for colored driving equally superposing three different $\omega_d$'s showing that broadband driving can induce suppression over a wide range of the band. The systematic variation of the suppression with $\alpha$ and $\gamma_l$ is shown in \fir{fig:stack_cooling}(c) where the relative phase fluctuations averaged over all modes in the interbilayer band are plotted. This indicates that optimal cooling occurs with a moderate coupling and weak intrinsic damping of the interbilayer plasmon. 

\begin{figure}[tb]
\centering
\includegraphics{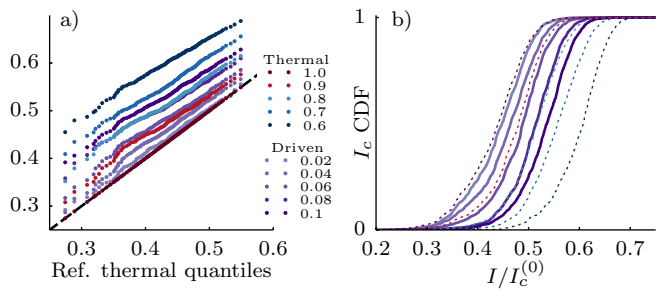}
\caption{(a)~Q-Q plot of switching current distributions for driving strengths $\eta_h = 0\rightarrow 0.1$, at $T/T_c \approx 0.7$. Thermal curves are relative to this temperature. The dashed line indicates the initial thermal distribution. 
(b)~Numerically computed CDF of switching current at $\alpha = 1$ with a sweep time $\Delta t = 1 $ ns. Solid lines are the driven stack with different driving strengths as in (a) while the dotted lines correspond to thermal benchmarks.}
\label{fig:stack_systematics_and_Ic}
\end{figure}

A complementary characterization of the stack is given by the $c$-axis superconducting transport properties quantified by the switching current distribution. This was obtained by sweeping in a time $\Delta t = 1$ ns the bias current $I(t)$ linearly in time from zero up to the critical current $I_c$. The tilt of the washboard potential of every junction in the stack increases until a phase slip event occurs, at which point the potential difference across the stack becomes finite. In \Fir{fig:stack_systematics_and_Ic}(a) the Q-Q plot for the computed statistics of this process are reported. This compares the quantiles of the original thermal distribution to those of the stack subjected to different driving strengths $\eta_h$. The curves indicate a shift in the mean of the switching distribution and a reduction in its spread, both of which are expected for a switching distribution at a lower temperature \cite[pp.~207--209]{Tinkham1996}.  

This tendency is confirmed in \Fir{fig:stack_systematics_and_Ic}(b) where the cumulative distribution function (CDF) for the switching current is plotted for the same set of drivings. The distribution is shifted to higher values of the bias current with increasing driving, analogous to the shift that is observed for thermal curves with lowering temperature. The suppression of phase fluctuations in the interbilayer band thus corresponds to a measurable cooling effect on an experimentally relevant figure of merit. 
Since $\hbar \omega_{h,i} > k_B T$ the intrabilayer modes remain superconducting with no phase slips induced.

\prlsection{Conclusion}
We have shown the suppression of phase fluctuations of Josephson plasmons in bilayer cuprates by selectively targeting an IR-active $c$-axis vibrational modes that modulate the Josephson plasma frequencies. At the difference frequency between intrabilayer and interbilayer plasma modes this driving can be exploited to implement parametric cooling. We have shown that a moderate capacitive coupling and low damping of the interbilayer modes is needed for this effect to be optimal. While the coupling in BSCCO-2212 is too weak, both YBCO and TBCCO-2201 satisfy these requirements making them strong candidate materials. Additionally they both possess phonon modes near the difference frequency with atomic motion in the intrabilayer junction. Related effects have already been observed in YBCO \cite{Mankowsky2014}, and the experiments reported in Refs.~\cite{Kaiser2014,Hu2014} may in fact rest on a related physical mechanism, despite the exact resonance condition for parametric cooling not strictly being met.

The proposed scheme may provide a pathway for dynamically stabilizing superconductivity above $T_c$ so long as superconducting coherence and a high frequency plasmon remains \cite{Dubroka2011}. For cooling to stay effective the interbilayer plasmon must remain underdamped even with increasing temperature. Future work includes extending the treatment to long junctions possessing phase fluctuations in the $ab$ plane \cite{Hoppner2014}, and investigating the possibility of dark-state cooling schemes \cite{Griessner2006} in the richer structure of tri-layer materials.

\prlsection{Acknowledgements}
This research is funded by the European Research Council under the European Union's Seventh Framework Programme (FP7/2007--2013)/ERC Grant Agreement no.~319286 Q-MAC
and resources were provided by the TNT project funded via EPSRC projects EP/K038311/1 and EP/J010529/1.

%


\pagebreak
\begin{widetext}
\begin{center}
\textbf{\large Supplemental Material:\\ Proposed parametric cooling of bilayer cuprate superconductors\\ by terahertz excitation}
\newline
\newline
S.~J.~Denny$^{1}$, S.~R.~Clark$^{1,2}$, Y.~Laplace$^{2}$, A.~Cavalleri$^{2,1}$, and D.~Jaksch$^{1,3}$
\small{
\newline
\emph{$^1$Clarendon Laboratory, University of Oxford, Parks Road, Oxford OX1 3PU, United Kingdom}
\newline
\emph{$^2$Max Planck Institute for the Structure and Dynamics of Matter, Hamburg, Germany}
\newline
\emph{$^3$Centre for Quantum Technologies, National University of Singapore, 3 Science Drive 2, Singapore 117543}
}
\end{center}
\end{widetext}

\renewcommand{\theequation}{S\arabic{equation}}
\renewcommand{\thefigure}{S\arabic{figure}}
\renewcommand{\thesection}{S\arabic{section}}
\renewcommand{\thepage}{S\arabic{page}}
\renewcommand{\bibnumfmt}[1]{[S#1]}
\renewcommand{\citenumfont}[1]{S#1}
\titlelabel{\thetitle:\quad}

\setcounter{equation}{0}
\setcounter{figure}{0}
\setcounter{table}{0}
\setcounter{page}{1}

\section{Electrodynamical model of a stack}
In the main text the electrodynamics of a superconducting stack is modeled  
using an approach developed by \textcite{sup_Machida2004}. Particular attention is paid in this approach to the 
penetration of the electric field into the superconducting sheets.
This is necessary to correctly treat the capacitive coupling between 
adjacent junctions, which is dominant in the short-junction regime.
In this section we extend their derivation to incorporate an effective 
time-dependent permittivity arising from the phonon driving.

To start we introduce some convenient notation. We will use the label $i$ for all properties relating to the 
superconducting layers. For the Josephson junction formed with the insulator between superconducting layers $(i,i-1)$
we instead use the label $I$ for all its properties. The framework of \textcite{sup_Machida2004} is semi-microscopic since it is 
based on the Schr\"odinger equation for the macroscopic superfluid wavefunction and 
derives the electric field coupling from the wavefunction dynamics.
This culminates in the equation 
\begin{equation}\label{eq:MS_gen_josephson}
   \frac{\Phi_0}{2\pi} \partial_t \phi_I =
     -\frac{\lambda_{c,i}^2}{\epsilon^s_i}\rho_i
     + \frac{\lambda_{c,i-1}^2}{\epsilon^s_{i-1}}\rho_{i-1}
     + \int_{z_{i-1}}^{z_i} E_z dz.
\end{equation}
for the Josephson phase dynamics of the gauge invariant phase $\phi_I$ between 
superconducting layers $i$ and $i-1$, where $\Phi_0$ is the flux quantum. This generalizes the 
conventional Josephson relation, given by the last term in \eqr{eq:MS_gen_josephson}, to include 
the contribution arising from the superconducting layer charge densities $\rho_i$, given the superconducting
layers have a dielectric constant $\epsilon^s_i$ and a $c$-axis Debye length $\lambda_{c,i}$.
By considering \cite{sup_Machida2004} the variation of the scalar, chemical and electrochemical 
potentials across a junction leads to the relation
\begin{equation}
   \rho_i = -\frac{ \epsilon^s_i}{\lambda_{c,i}^2}\left( {\Theta}_i + 
\frac{\Phi_0}{2\pi} \frac{\partial \phi_{I}}{\partial t} \right)
\end{equation}
where $\Theta_i$ is the scalar potential in the $i^\text{th}$ superconducting layer.
This, together with \eqr{eq:MS_gen_josephson} and Gauss' law
\begin{equation*}
   \partial_z E_z = \frac{\rho_i}{ \epsilon^s_i},
\end{equation*}
gives rise to an intuitive screening equation for the electric field 
inside the superconducting sheets,
\begin{equation}
   \partial_z^2 E_{z,i} = \frac{1}{\lambda_{c,i}^2} E_{z,i}.
\end{equation}
This acts in addition to the well-established London screening relation for the 
magnetic field inside the superconducting layers,
\begin{equation}
   \partial_z^2 B_{y,i} = \frac{1}{\lambda_{L,i}^2} B_{y,i},
\end{equation}
where $\lambda_{L,i}$ is the effective London penetration depth for 
the $i^\text{th}$ superconducting layer. For simplicity we will assume that $\phi_I$ is 
independent on $y$ in the plane and so only consider $x$ spatial phase variations.

For given $E$ and $B$ fields in the insulating layers, we now solve these 
equations to relate the electric field with the charge on the layer, and 
likewise the magnetic field with the screening currents.
At this point our derivation differs from that of 
\textcite{sup_Machida2004}. In particular we model the driven $c$-axis phonon mode
as a time-dependent modulation of the permittivity $\epsilon_I(t)$ about its equilibrium value $\epsilon_I$
for the insulator between the superconducting layers $i$ and $i-1$. Consequently we now
deal with the boundary conditions at the superconductor/insulator interface to 
properly take account of this effect. 

For a superconducting layer of thickness $t_i$, with boundary conditions 
of $E = \{E_d,\ E_u\}$ for its \emph{downside} (d) and \emph{upside} (u), 
corresponding to $z - z_0 = \{0,\ t_i\}$ respectively, we find that the 
electric field inside is given by
\begin{align*}
   E_{z,i}(z - z_0) &= E_d \cosh \left( \frac{z-z_0}{\lambda_{c,i}} 
\right) \\
                     &\quad+ \frac{E_u - E_d \cosh 
(t_i/\lambda_{c,i})}{\sinh (t_i/\lambda_{c,i})} \sinh \left( 
\frac{z-z_0}{\lambda_{c,i}} \right).
\end{align*}
There is a jump in the electric field between the insulator and the 
superconducting layer
\begin{equation*}
   E_I = \frac{\epsilon^s_i}{\epsilon_{I}(t)} E_d^i,
\end{equation*}
which arises from the boundary condition for the electric field at the 
interface between two media.
We assume there is no sheet charge at the surface so instead screening occurs 
due to a charge distribution over the Debye length scale.
This leads to
\begin{subequations}\label{eq:rho_updown}
\begin{align}
   \rho_i^d &= \frac{\epsilon_{I+1}(t)E_{I+1} - 
\epsilon_{I}(t)\cosh(t_i/\lambda_{c,i})E_{I}}{\lambda_{c,i} \sinh 
(t_i/\lambda_{c,i})},  \\
   \rho_i^u &= \frac{\epsilon_{I+1}(t)\cosh(t_i/\lambda_{c,i}) E_{I+1} 
- \epsilon_{I}(t)E_{I}}{\lambda_{c,i} \sinh (t_i/\lambda_{c,i})}
\end{align}
\end{subequations}
for the charge densities on the downside and upside of the $i^\text{th}$ 
superconducting layer.

\begin{figure}[tb]
   \centering
   \includegraphics{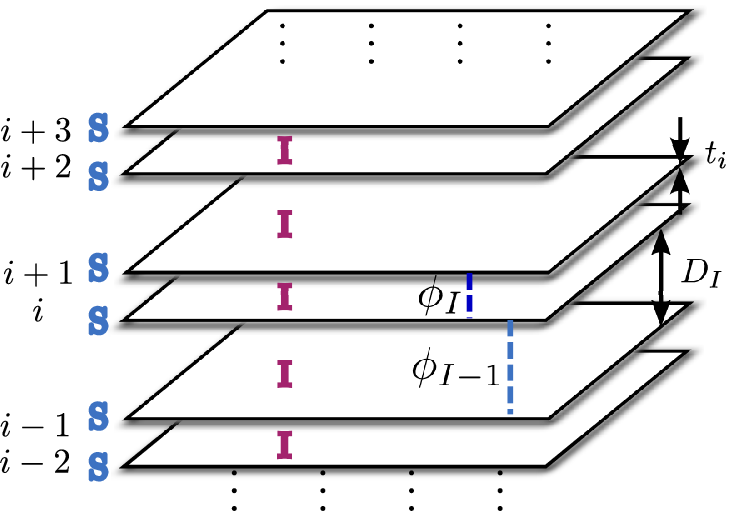}
   \caption{Schematic of the bilayer cuprate for the electrodynamical model.
   Superconducting sheets of finite thickness $t_i$ are separated by insulating regions of thickness $D_I$.
   The Josephson effect provides coupling between the superconducting sheets.}
   \label{fig:appendix_schematic}
\end{figure}

We now go back to the general Josephson relation 
\eqr{eq:MS_gen_josephson} and substitute for the charge densities evaluated in \eqr{eq:rho_updown} 
yielding
\begin{equation}\label{eq:Machida_phi_E}
   \frac{\Phi_0}{2\pi} \partial_t \phi_{I} =
       s_{I,I-1}^{C,d} E_{I-1} + D_{I}^C E_{I} + s_{I+1,I}^{C,u} E_{I+1},
\end{equation}
where the diagonal $D^C$ and off-diagonal $s^C$ capacitive couplings are
\begin{align}
   D_{I}^C(t) &= D_{I}
       + \frac{\epsilon_{I}(t)}{\epsilon^s_i} \lambda_{c,i} \coth 
\left(\frac{t_{i}}{\lambda_{c,i}}\right) \notag\\
       &\qquad+ \frac{\epsilon_{I}(t)}{\epsilon^s_{i-1}} 
\lambda_{c,i-1} \coth \left(\frac{t_{i-1}}{\lambda_{c,i-1}}\right), \notag\\
   s_{I+1,I}^{C,u}(t)       &= 
-\frac{\epsilon_{I+1}(t)}{\epsilon^s_i}\frac{\lambda_{c,i}}{\sinh 
(t_{i}/\lambda_{c,i})}, \notag\\
   s_{I,I-1}^{C,d}(t)       &= 
-\frac{\epsilon_{I-1}(t)}{\epsilon^s_{i-1}}\frac{\lambda_{c,i-1}}{\sinh 
(t_{i}/\lambda_{c,i-1})}. \label{eq:Machida_Dc_Sc}
\end{align}
Here $D_I$ is the thickness of the insulator for the $I^\textrm{th}$ junction. A similar equation relating the phase gradient to the magnetic field may 
be derived by solving for and eliminating the screening currents in the 
superconducting layer, giving
\begin{equation}\label{eq:Machida_phi_B}
     \frac{\Phi_0}{2\pi} \partial_x \phi_{I} =
       s_{I,I-1}^L B_{I-1} + D_{I}^L B_{I} + s_{I+1,I}^L B_{I+1}, 
\end{equation}
where the diagonal $D^L$ and off-diagonal $s^L$ inductive couplings are
\begin{align*}
   D_{I}^L &= D_{I} + 
\lambda_{L,i}\coth\left(\frac{t_i}{\lambda_{L,i}} \right) \\
                &\qquad\qquad+ 
\lambda_{L,i-1}\coth\left(\frac{t_{i-1}}{\lambda_{L,i-1}} \right), \\
   s_{I+1,I}^L &= - \frac{\lambda_{L,i}}{\sinh(t_i/\lambda_{L,i})}.
\end{align*}
We now summarize \eqr{eq:Machida_phi_E} and \eqr{eq:Machida_phi_B} in 
matrix form as
\begin{align}
   \frac{\Phi_0}{2\pi} \partial_t \vec \phi &= \mathbf{C}(t) \vec E, \label{eq:Machida_E}\\
   \frac{\Phi_0}{2\pi} \partial_x \vec \phi &= \mathbf{L} \vec B, \label{eq:Machida_B}
\end{align}
where $\mathbf{C}(t)$ and $\mathbf{L}$ contain the capacitive and inductive couplings, respectively, and with the
former being time-dependent owing to $\epsilon_{I}(t)$. 

Maxwell's equation for the insulating layers is given by
\begin{equation}
   \partial_x B_I^y = \frac{\epsilon_{I}(t)}{c^2} \partial_t E_I^z + \mu_0 j_I^z, \label{eq:maxwell_insulator}
\end{equation}
where $j^z_I = j^c_I \sin \phi_I + \sigma E^z_I$ is the $c$-axis current through the junction, composed of the Josephson supercurrent, quantified by its critical current $j^c_I$ 
and the quasi-particle current, quantified by its conductivity $\sigma$. Substituting \eqr{eq:Machida_E} and \eqr{eq:Machida_B} into \eqr{eq:maxwell_insulator} we reach a closed equation concerning just the phase dynamics,
\begin{multline}\label{eq:parametric_phi_EoM}
   \frac{\Phi_0}{2\pi} \sum_J \left[ \mathbf L^{-1} \right]_{IJ} 
\partial_x^2 \phi_J  =  \\
   \frac{\Phi_0}{2\pi} \frac{\epsilon_{I}(t)}{c^2} \sum_J \left[\mathbf C^{-1}(t) \right]_{IJ} \partial^2_t \phi_J  + \mu_0 j_I^c \sin(\phi_I)\\
    + \frac{\Phi_0}{2\pi c^2} \sum_J \left\{\epsilon_{I}(t)\left[\partial_t \mathbf C^{-1}(t) \right]_{IJ} + \frac{\sigma}{\epsilon_0} \left[ \mathbf C^{-1}(t) 
\right]_{IJ}\right\} \partial_t \phi_J.
\end{multline}
Aside from the assumptions of the layered 
stack and a time--dependent $\epsilon_I(t)$ within the insulating layers, this equation provides a 
general description of Josephson phase dynamics in the cuprates.
In particular each term in \eqr{eq:parametric_phi_EoM} has an 
intuitive interpretation. The $\sin(\phi_I)$ term describes 
supercurrents, the $\partial^2_x \phi_J$ accounts for inductive coupling, the $\partial^2_t \phi_J$
accounts for capacitive coupling, and the $\partial_t \phi_J$ terms describes quasiparticle 
current between layers.

As described in the main text we focus on short junctions where we can assume $\phi_I$ is $x$-independent. For the moment we also drop the 
incoherent quasiparticle current terms since its damping contribution to the dynamics are accounted for later in the main text once we move to a Langevin description.
For a given junction $I$ we are then left with
\begin{equation}\label{eq:MachidaSakaiParametric}
  \partial_t^2 \phi_I + \frac{2\pi}{\Phi_0} \sum_J  \frac{C_{IJ}(t) j_J^c}{\epsilon_0 \epsilon_{J}(t)} \sin(\phi_J) = 0.
\end{equation}
Writing out the individual terms in the sum we obtain
\begin{multline*}
  \partial_t^2 \phi_I +  \frac{2\pi}{\Phi_0\epsilon_0} \left[j_{I-1}^c\frac{s^{C,d}_{I,I-1}(t)}{\epsilon_{{I-1}}(t)} \sin(\phi_{I-1}) \right. \\
      +  j_I^c\frac{D^C_I(t)}{\epsilon_{I}(t)} \sin(\phi_I) \\
      + \left. j_{I+1}^c\frac{s^{C,u}_{I+1,I}(t)}{\epsilon_{{I+1}}(t)} \sin(\phi_{I+1})\right] = 0.
\end{multline*}
Note that from \eqr{eq:Machida_Dc_Sc} the ratios $s^{C,d}_{I,I-1}(t)/\epsilon_{I-1}(t)$ and $s^{C,u}_{I+1,I}(t)/\epsilon_{I+1}(t)$ describing off-diagonal capacitive couplings are in fact time-independent. Moreover in the diagonal coupling $D^{C}_I(t)/\epsilon_{I}(t)$ only the first term $D_{I}/\epsilon_{I}(t)$ retains the time-dependence and so it alone accounts for the modulation by the phonon.

We now make the assumption that all superconducting layers are identical (i.e.~$\epsilon^s_i = \epsilon^s_{i+1}=\epsilon^s$, $t_i = t_{i+1}=t$, $\lambda_{c,i} = \lambda_{c,i+1} = \lambda_c$), and temporarily neglect the explicit time-dependence in the parameters. We define a dimensionless coupling parameter $\alpha$ according to
\begin{equation*}
  \frac{\alpha^2}{(2\alpha + 1)^2} = \frac{s_{I,I-1}^{C,d}s_{I+1,I}^{C,d}}{D_I^C D_{I+1}^C}.
\end{equation*}
We can then define a frequency associated with the junction $I$ with
\begin{equation*}
  \varpi_I^2 = \frac{1}{2\alpha + 1} \frac{2\pi}{\Phi_0} \frac{j_I^c D_I^C}{\epsilon_0 \epsilon_I},
\end{equation*}
which enables us to rewrite the equation of motion \eqr{eq:MachidaSakaiParametric} as
\begin{multline*}
  \partial_t^2 \phi_I + \alpha \varpi_{I-1}^2 \sin(\phi_{I-1}) - (2\alpha+1)\varpi_I^2\sin(\phi_I) \\
    + \alpha\varpi_{I+1}^2 \sin(\phi_{I+1}) = 0.
\end{multline*}
The result is that the microscopic material parameters have been expressed as phenomenological frequencies $\varpi_I$ and a capacitive coupling constant $\alpha$.  
This form is particularly useful, since it coincides with that of the  well-known Koyama-Tachiki model \cite{sup_Koyama1996}. Note that the limit of $\alpha =0$, corresponding to extremely thick superconducting layers, yields uncoupled junctions. Experimentally the coupling strength $\alpha$ has been determined to take values in the range $0.1 \rightarrow 5$.

To connect with the driven equation of motion in the main text, we consider again \eqr{eq:MachidaSakaiParametric} without dropping the time-dependence, and find that the equation of motion is modified by the introduction of a relative driving strength
\begin{align*}
  \eta_I(t) &= (2\alpha+1) \Lambda_I \left( \frac{\epsilon_I}{\epsilon_I(t)} - 1 \right) \\
	   &= \eta_I \sin^2(\omega_d t).
\end{align*}
where $\Lambda_I = D_I/D_I^C$, allowing \eqr{eq:MachidaSakaiParametric} to be expressed in the form presented in the main text,
\begin{multline*}
  \partial_t^2 \phi_I + \alpha \varpi_{I-1}^2 \sin(\phi_{I-1}) - [2\alpha+1+\eta_I(t)]\varpi_I^2\sin(\phi_I) \\
   + \alpha\varpi_{I+1}^2 \sin(\phi_{I+1}) = 0,
\end{multline*}
again using $\alpha$ and $\varpi_I$ as defined above.

\section{Analysis: two junction special case}
In this section we present the detailed analysis of the two junction unit cell culminating in the derivation of expressions for the cooling limit and rate.

\subsection{Transformation to normal modes}
The equations of motion for the special case of two junctions is 
\begin{equation*}
  \begin{gathered}
    \partial_t^2 \phi_l - [2\alpha + 1 + \eta_l(t)] \varpi_l^2 \sin(\phi_l) + \alpha \varpi_h^2 \sin(\phi_h) = 0, \\
    \partial_t^2 \phi_h + \alpha \varpi_l^2 \sin(\phi_l) - [2\alpha + 1 + \eta_h(t)] \varpi_h^2 \sin(\phi_h) = 0.
  \end{gathered}
\end{equation*}
We linearize about the unbiased equilibrium $\phi_i = 0$, with $i=\{l,h\}$, to get
\begin{equation*}
  \begin{gathered}
    \partial_t^2 \phi_l - [2\alpha + 1 + \eta_l(t)] \varpi_l^2 \phi_l + \alpha \varpi_h^2 \phi_h = 0, \\
    \partial_t^2 \phi_h + \alpha \varpi_l^2 \phi_l - [2\alpha + 1 + \eta_h(t)] \varpi_h^2 \phi_h = 0,
  \end{gathered}
\end{equation*}
and expand the driving perturbations $\eta_i(t)$ into static and dynamic parts, i.e.
\begin{align*}
  \eta_i(t) &= \eta_i \sin^2(\omega_d t)\\
            &= \half\eta_i \left[ 1 - \cos(2\omega_d t)\right].
\end{align*}
We diagonalize the dynamics
\begin{equation*}
  \partial_t^2 \left( \begin{array}{c}
                       \phi_l \\
                       \phi_h
                      \end{array} \right)
  = \left( \begin{array}{cc}
            [2\alpha + 1 + \half \eta_l]\varpi_l^2 & -\alpha \varpi_h^2 \\
            -\alpha \varpi_l^2 & [2\alpha + 1 + \half \eta_h]\varpi_h^2
           \end{array} \right)
    \left( \begin{array}{c}
                       \phi_l \\
                       \phi_h
                      \end{array} \right)
\end{equation*}
to define a \emph{fixed} transformation $\mathbf P$ from ($\phi_l$,$\phi_h$) to the normal coordinates of the phase, $\varphi_l$ and $\varphi_h$,
\begin{equation*}
  \left( \begin{array}{c}
    \varphi_l \\
    \varphi_h
  \end{array} \right)
  = \mathbf{P}
    \left( \begin{array}{c}
    \phi_l \\
    \phi_h
  \end{array} \right).
\end{equation*}
The normal coordinates $\varphi_l$ and $\varphi_h$ are then associated to normal frequencies $\omega_l$ and $\omega_h$ which are driving dependent.
The equations of motion for the normal coordinates are then
\begin{gather*}
  \partial_t^2 \varphi_l + \omega_l^2(t) \varphi_l + \Delta_h(t) \varphi_h = 0, \\
  \partial_t^2 \varphi_h + \omega_h^2(t) \varphi_h + \Delta_l(t) \varphi_l = 0.
\end{gather*}
The time-dependent character of $\eta_i(t)$ is reflected in the time-dependence of the parameters $\omega_i(t)$ and $\Delta_i(t)$.  
The frequencies $\omega_i(t)$ modulate in time as
\begin{align*}
  \omega_l^2(t) &= \omega_l^2 - \half\left[\frac{\alpha^2}{(2\alpha + 1)^2} \eta_h + \eta_l\right] \varpi_h^2 \cos (2\omega_d t)r^2 + \mathcal{O}(r^4), \\
  \omega_h^2(t) &= \omega_h^2 - \half \eta_h \varpi _h^2 \cos(2\omega_d t) + \mathcal{O}(r^2),
\end{align*}
and the off-diagonal terms (the couplings) are derived as
\begin{align*} 
    \Delta_l(t) &= -\half\eta_h  \varpi_h^2 \frac{\alpha^2}{(1+2\alpha)^2} \cos (2\omega_d t) \ r^2 + \mathcal{O}(r^4), \\
    \Delta_h(t) &= -\half\eta_h  \varpi_h^2 \cos (2\omega_d t)  + \mathcal{O}(r^2),
\end{align*}
after expanding in the small parameter $r = \varpi_l/\varpi_h$.

\subsection{Approximate coupled oscillators}
Our approximation is that we neglect the time-dependence of $\omega_{i}(t)$, and assume that only the modulated coupling terms in this frame give rise to cooling. We write down a Hamiltonian for these normal coordinates in terms of $\varphi_i$, and its conjugate momentum $p_i$ \footnote{As the canonical momentum $p_i$ is conjugate to a normal mode of the gauge-invariant phase difference of the junction, there is no convenient physical interpretation to be made regarding this quantity. In particular, it is not simply related to number differences between the superconducting sheets.},
\begin{multline}\label{eq:effective_twoJJ_H_sup}
 H = \frac{p_l^2}{2\Delta_l}
     + \frac{p_h^2}{2\Delta_h}
     + \half \Delta_l \omega_l^2 \varphi_l^2
     + \half \Delta_h \omega_h^2 \varphi_h^2 \\
     + \cos(2\omega_d t)\  \Delta_l \Delta_h \varphi_l \varphi_h,
\end{multline}
in which we have separated out the time dependence of $\Delta_i(t) = \Delta_i \cos (2\omega_d t)$.
To exploit the parametric cooling results of \textcite{sup_Vyatchanin1977} we quantize and apply a rotating wave approximation.
This amounts to promoting the phase normal coordinates $\varphi_i$ and their conjugate momenta $p_i$ to operator status,
\begin{align*}
  \varphi_i &\rightarrow \hat \varphi_i \\
  p_i &\rightarrow \hat p_i,
\end{align*}
and defining ladder operators according to
\begin{align*}
  \hat a &= \sqrt{\frac{\Delta_l \omega_l}{2\hbar}} \left( \hat \varphi_l + \frac{i}{\Delta_l \omega_l}\hat p_l \right), \\
  \hat a^\dagger &= \sqrt{\frac{\Delta_l \omega_l}{2\hbar}} \left( \hat \varphi_l - \frac{i}{\Delta_l \omega_l}\hat p_l \right), \\
  \hat b &= \sqrt{\frac{\Delta_h \omega_h}{2\hbar}} \left( \hat \varphi_h + \frac{i}{\Delta_h \omega_h}\hat p_h \right), \\
  \hat b^\dagger &= \sqrt{\frac{\Delta_h \omega_h}{2\hbar}} \left( \hat \varphi_h - \frac{i}{\Delta_h \omega_h}\hat p_h \right),
\end{align*}
which after dropping the zero-point terms yields
\begin{eqnarray}\label{eq:Vyatchanin_H_preRWA}
  \hat{H} &=& \hbar \omega_l \hat a^\dagger \hat a + \hbar \omega_h \hat b^\dagger \hat b \nonumber \\
  &&  + ~ \hbar \sqrt{\frac{\Delta_l\Delta_h}{\omega_l \omega_h}} \cos(2\omega_d t) \left( \hat a + \hat a^\dagger\right)\left( \hat b + \hat b^\dagger\right).
\end{eqnarray}
In this system the frequencies are such that $\omega_l < \omega_h$, so we make the rotating wave approximation, i.e.~dropping the counter-rotating terms $\hat a \hat b$ and $\hat a^\dagger \hat b^\dagger$, giving the Hamiltonian
\begin{equation}\label{eq:Vyatchanin_H}
  \hat{H} = \hbar \omega_l \hat a^\dagger \hat a + \hbar \omega_h \hat b^\dagger \hat b + \hbar \left[ \kappa(t) \hat a \hat b^\dagger +  \kappa^*(t) \hat a^\dagger \hat b\right].
\end{equation}
In this approximation we thus end up with a coupling modulated according to $\kappa(t) = \kappa_0 \exp(2i\omega_d t)$, and $\omega_d \approx \omega_h - \omega_l$, $\kappa_0 \ll \omega_l$.
More explicitly the magnitude of the coupling $\kappa_0$ is
\begin{gather*}
  \kappa_0^2 = \frac{\eta_h^2 \omega_h^2}{4} g(\alpha)\ r + \mathcal{O}(r^3), \\
  g(\alpha) = \frac{\alpha ^2 }{(2 \alpha +1)^2 \sqrt{3 \alpha ^2+4 \alpha +1}},
\end{gather*}
The function $g(\alpha)$ is maximised at $\alpha = \alpha_0 \approx 1.07$, at which $g(\alpha_0) \approx 3.9 \times 10^{-2}$.

\section{Resonant energy exchange}
The physics underlying the parametric cooling implemented by this modulated coupling can be exposed in the Hamiltonian formulation, by means of a unitary transformation.
If we consider the Hamiltonian \eqr{eq:Vyatchanin_H} and transform into a rotating frame with unitary
\begin{equation*}
  \hat U = \exp \left( \frac{-i\hat A t}{\hbar} \right), \quad \hat A = -2\hbar \omega_d \hat a^\dagger a,
\end{equation*}
we arrive at a very similar Hamiltonian,
\begin{equation}
  \hat H = \hbar (\omega_h + \Delta \omega) \hat a^\dagger \hat a + \hbar \omega_h \hat b^\dagger \hat b + \hbar \left(\kappa_0 \hat a^\dagger \hat b + \kappa^*_0 \hat a \hat b^\dagger \right). \nonumber
\end{equation}
The time dependence of the coupling has been removed, and the oscillators now formally appear in this frame to have the same frequency up to a detuning $\Delta \omega = 2\omega_d - (\omega_h - \omega_l)$. Thus parametric modulation allows resonant energy exchange between the oscillators via down- and up-conversion of quanta.

\subsection{Quantum master equation}
We include damping and noise due to the coupling to long-wavelength phonon modes and quasi-particle currents by coupling each oscillator to its own Markovian reservoir.  
The relaxation times are $\gamma_l^{-1}$ and $\gamma_h^{-1}$ for the interbilayer and intrabilayer modes respectively, and their equilibrium oscillator occupation numbers are $\nu_l$ and $\nu_h$. The dissipative system dynamics are described by a master equation for the density matrix,
\begin{align}
  \frac{\partial \hat \rho}{\partial t} = &-\gamma_l 
      (1 + \nu_l)(\hat a^\dagger \hat a \hat \rho - 2 \hat a \hat \rho \hat a^\dagger + \hat \rho \hat a^\dagger \hat a) \notag\\
      &-\gamma_l \nu_l (\hat a \hat a^\dagger \hat \rho - 2 \hat a^\dagger \hat \rho \hat a + \hat \rho \hat a \hat a^\dagger)  \notag\\
      &-\gamma_h(1 + \nu_h)(\hat b^\dagger \hat b \hat \rho - 2 \hat b \hat \rho \hat b^\dagger + \hat \rho \hat b^\dagger \hat b) \notag\\
      &-\gamma_h \nu_h (\hat b \hat b^\dagger \hat \rho - 2 \hat b^\dagger \hat \rho \hat b + \hat \rho \hat b \hat b^\dagger)  \notag\\
      &+\frac{1}{i\hbar} \commutator{\hat H}{\hat \rho}.\label{eq:master_equation}
\end{align}
Notice that the transformation $\hat U$ does not affect the noise processes in this equation. The parametric coupling causes the oscillators to resonantly exchange quanta, with the upward transition rate proportional to the number of quanta in the interbilayer oscillator, and likewise for the downward rate. Crucially, when the oscillators are at the same temperature, the interbilayer oscillator contains more quanta on average, and so there is a net flow $\varphi_l \rightarrow \varphi_h$. The driven system will therefore reach a steady state with the $\varphi_l$ interbilayer oscillator containing fewer quanta on average than its thermal distribution for temperature $T$ would produce and so it is cooled. Correspondingly the $\varphi_h$ intrabilayer oscillator is heated.

This system can be solved for the steady state \cite{sup_Vyatchanin1977}. Specifically, from the above master equation, we produce a set of equations for the time evolution of the second-order moments,
\begin{equation}
  \frac{d}{dt} \left( \begin{array}{c}
                       \expt{\hat{a}^\dagger \hat{a}} \\ \expt{\hat{b}^\dagger \hat{b}} \\ \expt{\hat{a}^\dagger \hat{b}} \\ \expt{\hat{a} \hat{b}^\dagger}
                      \end{array}\right)
  =
          \mathbf{M}
         \left( \begin{array}{c}
          \expt{\hat{a}^\dagger \hat{a}} \\ \expt{\hat{b}^\dagger \hat{b}} \\ \expt{\hat{a}^\dagger \hat{b}} \\ \expt{\hat{a}\hat{b}^\dagger}
          \end{array}\right)
          +\left(\begin{array}{c}
          \gamma_l \nu_l \\ \gamma_h \nu_h \\ 0 \\ 0
          \end{array}\right),
\end{equation}
and find that they form a closed system.
Above, $\mathbf{M}$ is defined by
\begin{widetext}
\begin{equation*}
  \mathbf{M} = \left( \begin{array}{cccc}
          -\gamma_l & 0 & \frac{i \kappa_0}{2} e^{-2i \omega_d t} & -\frac{i \kappa_0}{2} e^{2i \omega_d t} \\
          0 & -\gamma_h & -\frac{i \kappa_0}{2} e^{-2i \omega_d t} & \frac{i \kappa_0}{2} e^{2i \omega_d t} \\
          \frac{i \kappa_0}{2} e^{2i \omega_d t} & -\frac{i \kappa_0}{2} e^{2i \omega_d t} & -\frac{1}{2}(\gamma_l+\gamma_h) + i(\omega_h-\omega_l) & 0 \\
          -\frac{i \kappa_0}{2} e^{-2i \omega_d t} & \frac{i \kappa_0}{2} e^{-2i \omega_d t} & 0 & -\frac{1}{2}(\gamma_l+\gamma_h) - i(\omega_h-\omega_l)
         \end{array} \right).
\end{equation*}
We transform (e.g.) $\hat{b} \rightarrow \hat{b}\ e^{2i\omega_d t}$,  and reach
\begin{equation}
  \frac{d}{dt} \vec u
  =
          \mathbf{M}' \vec u
          +\vec v,
\end{equation}
now with
\begin{equation*}
 \vec u = \left( \begin{array}{c}
                       \expt{\hat{a}^\dagger \hat{a}} \\ \expt{\hat{b}^\dagger \hat{b}} \\ \expt{\hat{a}^\dagger \hat{b}} \\ \expt{\hat{a} \hat{b}^\dagger}
                      \end{array}\right), \qquad
 \vec v = \left(\begin{array}{c}
          \gamma_l \nu_l \\ \gamma_h \nu_h \\ 0 \\ 0
          \end{array}\right),
\qquad
  \mathbf{M}' = \left( \begin{array}{cccc}
          -\gamma_l & 0 & \frac{i \kappa_0}{2}  & -\frac{i \kappa_0}{2}  \\
          0 & -\gamma_h & -\frac{i \kappa_0}{2}  & \frac{i \kappa_0}{2}  \\
          \frac{i \kappa_0}{2}  & -\frac{i \kappa_0}{2}  & -\frac{1}{2}(\gamma_l+\gamma_h) - i\Delta \omega & 0 \\
          -\frac{i \kappa_0}{2}  & \frac{i \kappa_0}{2}  & 0 & -\frac{1}{2}(\gamma_l+\gamma_h) + i \Delta \omega
         \end{array} \right).
\end{equation*}
\end{widetext}
Formally, $\vec u = -\mathbf {M'}^{-1} \vec v$ yields the steady-state. For the interbilayer mode $\varphi_l$ oscillator this yields a steady-state occupation as
\begin{equation}
  \expt{\hat{a}^\dagger \hat{a}}_\text{ss} = \nu_l - \mathcal{S}(\nu_l - \nu_h),
\end{equation}
where the scale factor $0 \leq \mathcal{S} \leq 1$ is defined as $\mathcal{S} = \zeta/\gamma_l\chi$ with
\begin{align*}
  \zeta &= \frac{\kappa_0^2 (\gamma_l + \gamma_h )}{\Delta \omega^2 + (\gamma_l + \gamma_h)^2}, \\
  \chi &= 1 + \frac{\zeta(\gamma_l + \gamma_h)}{\gamma_l \gamma_h}, \\
  \Delta \omega &= 2\omega_d - (\omega_h - \omega_l),
\end{align*}
as stated in the main text. It is useful to consider some limits of this result. 
First, if there is either no coupling $\kappa_0=0$ between the plasmon modes, or no damping $\gamma_h=0$ on the intrabilayer mode $\varphi_h$, then $\mathcal{S} =0$ implying that there is no cooling. 
Second, in the limit of an undamped interbilayer mode $\gamma_l \rightarrow 0$ then $\mathcal{S}=1$ and the suppression of the interbilayer's occupation reaches the minimum of $\expt{\hat{a}^\dagger \hat{a}}_\text{ss} = \nu_h$ set by the intrabilayer mode. 
Third, assuming that $\gamma_h>0, \kappa_0>0$ we find that $\mathcal{S}$ decreases monotonically with $\gamma_l$. 
We see also that the interbilayer mode should be underdamped, as follows. 
We have $\kappa_0 \lesssim \omega_l$ as required by any model of coupled oscillators, and also $\gamma_l \lesssim \kappa_0$ in order that there is a significant cooling effect. 
It follows that $\gamma_l \lesssim \omega_l$, and so our theory requires that the oscillator be underdamped.
Finally, $\mathcal{S}$ exhibits the expected resonance around $\Delta\omega =0$.

To extract the cooling rate, we examine the eigenvalues of the matrix $\mathbf M'$.  
This is trivial numerically, but we gain insight from analyzing the perturbation of the eigenvalues of $\mathbf M'$ with $\kappa_0$.
In particular, in the limit of $\kappa_0 \ll (\gamma_l,\gamma_h)$ we use second-order perturbation theory to estimate the rate.   
We decompose $\mathbf M' = \mathbf M_0 + \delta \mathbf{M}$, with $\mathbf M_0 = \diag[-\gamma_l,-\gamma_h,-\half(\gamma_l+\gamma_h),-\half(\gamma_l+\gamma_h)]$ and
\begin{equation*}
    \mathbf{\delta M} = \frac{i\kappa_0}{2}\left( \begin{array}{cccc}
          \cdot & \cdot & +1  & -1  \\
          \cdot & \cdot & -1  & +1  \\
          +1  & -1  & \cdot & \cdot \\
          -1  & +1  & \cdot & \cdot
         \end{array} \right).
\end{equation*}
Applying standard perturbation theory results in a second order correction to the slowest eigenmode,
\begin{equation*}
  \gamma_\text{dr} = \gamma_l + \frac{\kappa_0^2}{\gamma_h - \gamma_l} + \mathcal{O}(\kappa_0^3).
\end{equation*}
We see that the cooling rate is determined by the thermalization timescale of the interbilayer mode, somewhat accelerated by the driving term. This indicates the trade-off between the cooling having a reduced $\mathcal{S}$ due to $\gamma_l>0$ verses a faster cooling rate when the interbilayer has some dissipation. For the parameters shown in the main text \fir{fig:two_junctions_phase_cooling}, this yields a rate $\gamma_\text{dr} = 5.62 \times 10^{10} \text{ s}^{-1}$, as opposed to $5.75 \times 10^{10} \text{ s}^{-1}$ resulting from a numerical evaluation of the eigenvalues.  
This all depends on the decay rates $\gamma_l$ and $\gamma_h$ differing sufficiently that degenerate perturbation theory is not required. In practice this requirement is satisfied, as $\gamma_h > \gamma_l$ for systems of interest here.  

To summarize, this section presents a theoretical description of the cooling process as a transformation in which the two normal mode oscillators are brought into resonance.  
In this picture, the parametric driving produces a linear coupling dependent on the driving magnitude, which accurately predicts both the magnitude and rate of cooling.

\section{Quantum calculation}\label{sec:quantum_calculation}
\begin{figure}[tbp]
  \centering
  \includegraphics{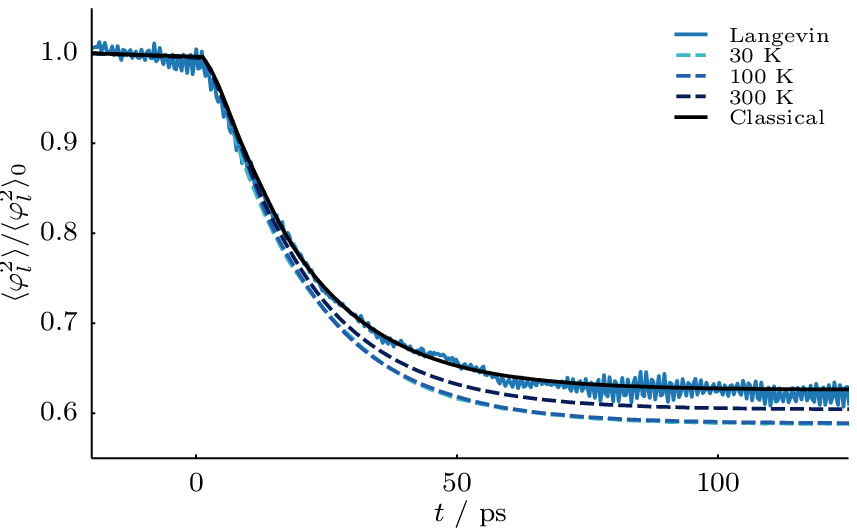}
  \caption{(a)~Comparison of the quantum calculation of \secr{sec:quantum_calculation} with the classical Langevin results presented in the main text.  
  `Langevin' presents the statistics from classical Langevin trajectories as in the main text, \fir{fig:two_junctions_phase_cooling}.  
  The dotted curves present quantum master equation predictions for $\expt{\varphi_l^2}$ for varying temperature $T$.  
  The quantity plotted is $(\expt{\varphi_l^2} - \expt{\varphi_l^2}_\text{qu}) / (\expt{\varphi_l^2}_0 - \expt{\varphi_l^2}_{\text{qu},0})$, namely the cooling of the \emph{classical fluctuations} of the interlayer plasmon.  
  We find the classical Langevin approach accurately describes the cooling of the thermal fluctuations over the entire relevant temperature range, with the quantum calculations in fact predicting enhanced cooling due to quenching of the intralayer mode.}
  \label{fig:quantum_pq_mod1}
\end{figure}
In this section we examine the approximation made in using a classical Langevin approach where quantum fluctuations are neglected.  
For the intrabilayer plasmon, even at room temperature we have a mean occupation $n = [\exp(\beta \omega_h) - 1]^{-1} \sim 0.25$ at 10 THz, and so it is a possibility that the quantum noise on this oscillator may have a measurable effect.  
We take \eqr{eq:effective_twoJJ_H_sup} and keep the time-dependence of the oscillator frequencies $\omega_i(t)$ such that $\hat H_l$ and $\hat H_h$ become time-dependent.
We evolve the system of equations for $\expt{\hat a^2}$, $\expt{\hat a^\dagger \hat a}$, \ldots{} that result from the master equation \eqr{eq:master_equation}, via 
\begin{equation}
\partial_t \expt{\hat O} = \Tr \left( \hat O\  \partial_t \hat \rho \right).
\end{equation}
This is more general than Vyatchanin calculation by (i) retaining counter-rotating terms, and (ii) keeping the time-dependence of the oscillator frequencies.

We are concerned with the cooling of {\em thermal} fluctuations of the interbilayer mode, and so we consider the figure of merit to be the fluctuations of the $\varphi_l$- and $p_l$-quadratures once the {\em quantum} component has been subtracted. In \fir{fig:quantum_pq_mod1}, we plot the evolution of $(\expt{\varphi_l^2} - \expt{\varphi_l^2}_\text{qu}) / (\expt{\varphi_l^2}_0 - \expt{\varphi_l^2}_{\text{qu},0})$ for a range of temperatures $T$  spanning a region around occupation numbers $n_l \sim 1$. Despite slight technical differences in the thermalization processes, there is excellent agreement between the classical Langevin calculation and the quantum master equation, with the quantum calculations indicating that the cooling of thermal fluctuations may be enhanced due to quenching of the interbilayer mode. The curve labelled `Classical' in \fir{fig:quantum_pq_mod1} is the (non-physical) high-temperature limit of the quantum master equation calculation, and as expected shows that the two approaches converge once both oscillators are in the limit of containing many quanta. The classical Langevin calculation thus accurately captures the cooling of the thermal fluctuations allowing the analysis in the main text to use classical methods for the non-linear regime of the stack.

\section{Temperature dependence of Josephson stack dynamics}
\begin{figure}[b!]
   \centering
   \includegraphics{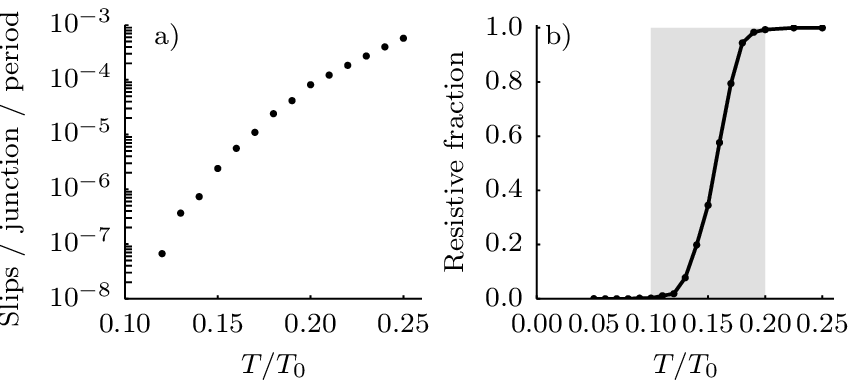}
   \caption{Temperature dependence of Josephson stack dynamics.  
   (a) Count of the number of phase slips in a 100-junction unbiased Josephson stack, normalised by the number of junctions and Josephson plasma timescale.
   (b) Fraction of trajectories that become resistive within a simulation time of 5 ns.  
   To the left of the shaded region, all trajectories remain superconducting, while to its right, all trajectories rapidly become resistive.}
   \label{fig:stack_critical_temperature}
\end{figure}

In this section we relate the temperature scale $T_0$ referenced in the main text to the critical temperature of a stack. At sufficiently low temperatures, the thermal fluctuations of the Josephson stack explore only the harmonic region of the washboard potential, while above a threshold temperature there is sufficient energy for spontaneous thermal phase slips to occur. If current-biased, the stack acquires a finite potential difference and thus becomes resistive at this point.  

In \fir{fig:stack_critical_temperature}(a) we consider a stack of 100 junctions with parameters as in the main text. Across the range of temperatures $0.1 \le T/T_0 \le 0.25$, after the thermalization burn-in we count the number of phase slips occurring in the stack, and normalise by the junction count and the Josephson period. Below $T / T_0 = 0.12$, zero phase slips occurred in the simulation time. 

In \fir{fig:stack_critical_temperature}(b) we apply a small bias $I/I_c = 0.1$ and record the fraction of trajectories which have become resistive after a simulation time of 5 ns (comparable to the switching current calculations presented in the main text \fir{fig:stack_systematics_and_Ic}, of $\Delta t = $ 1 ns). We classify trajectories as superconducting or resistive by comparing whether the center-of-mass phase is more than $\Delta \phi = 4\pi$ away from its value $\Delta t = $ 20 ps ago. This simple heuristic is capable of capturing the ``running'' state effectively, and is not sensitive to modest variation of the parameters $\Delta \phi$ and $\Delta t$ used. A region around $T / T_0 \sim 0.15$ is identified, which marks the onset of substantial thermal phase slipping behavior. Thus by simulating our proposed scheme at $T / T_0 = 0.1$, we argue that we are at a sufficiently high temperature to avoid the need for a full quantum treatment, while remaining below this transitional temperature range. 

%


\end{document}